# A Construction of Systematic MDS Codes with Minimum Repair Bandwidth

Yunnan Wu

*Abstract*—In a distributed storage system based on erasure coding, an important problem is the *repair problem*: If a node storing a coded piece fails, in order to maintain the same level of reliability, we need to create a new encoded piece and store it at a new node. This paper presents a construction of systematic $(n,k)$-MDS codes for $2k \le n$ that achieves the minimum repair bandwidth when repairing from $k+1$ nodes.

## I. INTRODUCTION

It is well known that erasure coding can be used to effectively provide reliability against node failures in a data storage system. For instance, we can divide a file of size $B$ into $k$ pieces, each of size $B/k$, encode them into $n$ coded pieces using an $(n,k)$ maximum distance separable (MDS) code, and store them at $n$ nodes. Then, the original file can be recovered from any set of $k$ coded pieces. This is optimal in terms of the redundancy–reliability tradeoff because $k$ pieces, each of size $B/k$, provide the minimum data for recovering the file, which is of size $B$.

One of the challenges for erasure coding–based distributed storage is the *repair problem* (introduced in [1]): If a node storing a coded piece fails or leaves the system, in order to maintain the same level of reliability, we need to create a new encoded piece and store it at a new node. If the source file is not available in the system (e.g., in an archival application), then the repair has to be done by accessing other encoded data only. A straightforward way to repair a failed node in a system based on $(n,k)$-MDS code is to let the new node download $k$ encoded pieces from a subset of the surviving nodes, reconstruct the original file, and compute the needed new coded piece. In this process, the new node incurred a network traffic of $k \times B/k = B$. Since network bandwidth could be a critical resource in distributed storage systems, an important consideration is to conserve the repair network bandwidth.

The repair problem amounts to the partial recovery of the code, whereas conventional erasure code design focused on the complete recovery of the information from a subset of the coded pieces. The consideration of the repair network traffic gives rise to new design challenges for erasure codes. This problem and its variants have been studied in recent years and various code constructions have been proposed. Next we briefly review the related existing work on the construction of erasure codes with reduced repair bandwidth.

In this paper, we focus on $(n,k)$-MDS codes, because they achieve the optimal reliability–storage tradeoff. Via a cut-based analysis, Dimakis et al. [1] presented a lower bound on the network bandwidth needed to repair one node in an $(n,k)$-MDS code. Under a symmetric setup where the replacement node downloads the same number of bits from each of $d$ nodes, it is shown that the total repair traffic has to be at least $\frac{Bd}{k(d-k+1)}$. The same bound for total repair traffic in fact also holds even if we relax the symmetric setup; this will be explained in Section II.

The cut lower bound on total repair traffic has been shown in [1]–[4] to be achievable using network coding, if we adopt a relaxed notion of repair — *function repair*, where the repaired code continues to be $(n,k)$-MDS but it may be different from the original code before the repair. However, it is not clear that this network coding scheme can be made always systematic (i.e., one copy of the data exists in uncoded form). From a practical standpoint, it is highly desirable to have the systematic feature, so that in normal cases, data can be read directly from the uncoded copy, without performing decoding.

Motivated in part by the pursuit of a systematic code with reduced repair bandwidth, in [5], Wu and Dimakis formulated a variant of the repair problem, called the *exact repair* problem, where the same code is always maintained before and after the repair. For the exact repair problem, [5] presented an *interference alignment* scheme and a vector version of it. The interference alignment scheme can achieve the cut bound $\frac{Bd}{k(d-k+1)}$ for $(n,2)$-MDS and the resulting code is systematic. However, the scheme cannot achieve the cut bound for general $k$.

Functional repair and exact repair are not the only possible models. In a recent work, Rashmi K.V. *et al.* [6] proposed a code construction that can achieve the cut bound for $d=k+1$. The construction of [6] essentially implements a hybrid functional and exact repair model. In the scheme, each node stores 2 symbols, $\boldsymbol{y}^T\boldsymbol{u}_i$ and $\boldsymbol{y}^T\boldsymbol{v}_i + \boldsymbol{z}^T\boldsymbol{u}_i$, where the $2k$ original information symbols are represented by two vectors $\boldsymbol{y} \in \mathbb{F}^k$ and $\boldsymbol{z} \in \mathbb{F}^k$. The vectors $\{\boldsymbol{u}_i\}$ can be chosen as the $n$ code vectors of an $(n,k)$-MDS code. If node $i$ fails, the first symbol $\boldsymbol{y}^T\boldsymbol{u}_i$ is exactly reconstructed; the second symbol $\boldsymbol{y}^T\boldsymbol{v}_i + \boldsymbol{z}^T\boldsymbol{u}_i$ is repaired to a new symbol that has the same form $\boldsymbol{y}^T * + \boldsymbol{z}^T\boldsymbol{u}_i$. Since $\{\boldsymbol{u}_i\}$ can be chosen based on any $(n,k)$-MDS code, we can in particular use a systematic $(n,k)$-MDS code. Thus, the code can expose half of the information symbols, $\boldsymbol{y}$, in uncoded form.

Having explained several repair models, we now reflect on the practical needs again. Both the MDS feature and the systematic feature are highly desirable in practice. However, providing the systematic feature does not necessarily require all symbols be exactly reconstructed. This motivates us to explore one avenue – Look for a systematic MDS code with a hybrid functional and exact repair model, where the systematic

Yunnan Wu is with Microsoft Research, One Microsoft Way, Redmond, WA, 98052. yunnanwu@microsoft.com.

symbols are exactly reconstructed and the nonsystematic symbols follow a functional repair model. Heading this direction, in this paper we present a construction of $(n,k)$-MDS codes for $2k \leq n$ that achieves the minimum repair bandwidth when repairing from $k+1$ nodes.

## II. REVIEW: CUTSET BOUND ON TOTAL REPAIR TRAFFIC

In this section we describe the cut bound for total repair traffic. The analysis amounts to a slight extension of the analysis in [1], [2]. Specifically, in [1], [2] the replacement node downloads the same number of bits from each of $d$ nodes; in the following Lemma 1, the replacement node is allowed to download any number of bits from the each of $d$ nodes. The same bound on total network traffic still holds.

**Lemma 1:** Consider $B$ bits being stored via an $(n,k)$-MDS code at $n$ nodes, where each node stores $\alpha = B/k$ bits. To repair any failed storage node by accessing $d \geq k$ nodes, the total incurred network traffic is at least $\frac{Bd}{k(d-k+1)}$.

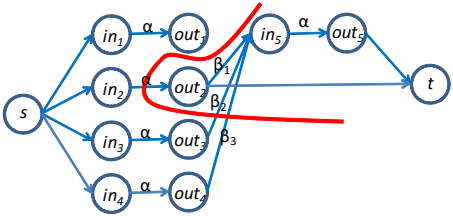

Fig. 1. Illustration of the proof of Lemma 1.

**Proof:** As in [1], [2], we consider the information flow graph that describes the repair problem as a network communication problem. The information flow graph is illustrated in Figure 1. In this graph, each storage node is represented by a pair of nodes, say $in_i$ and $out_i$, connected by an edge whose capacity is $\alpha$, the storage capacity of the node. There is a source node, $s$, which has the entire file. The source has infinite capacity edges to the $n$ storage nodes before the repair. In Figure 1, storage node 1 fails and we create a new storage node, node 5, which downloads $\beta_i$ bits from each of the three surviving nodes and then stores $\alpha$ bits; this is represented in Figure 1 by the edges $out_2 in_5$, $out_3 in_5$, and $out_4 in_5$ that enter node 5. There are also *data collectors*, each corresponding to one request to reconstruct the original data from a subset of the nodes. For example, the data collector $t$ in Figure 1 has infinite capacity edges from nodes 2 and 5, modeling that the file needs to be reconstructed by accessing storage nodes 2 and 5. By analyzing the cut between $s$ and the data collectors in the information flow graph, we can obtain bounds on the repair traffic. In particular, if the minimum cut between $s$ and a data collector $t$ is less than the size of the file, then we can conclude that it is impossible to reconstruct the file, regardless of what code we use. In the following, we use this cut argument to establish a bound on the total network traffic.

Without loss of generality, suppose the first storage node fails and node $n+1$ recovers the content stored at node 1 by downloading $\beta_i$ bits from node $i+1$ for $i = 2, \ldots, d+1$. Consider a data collector $t$ that connects to node $n+1$ and a

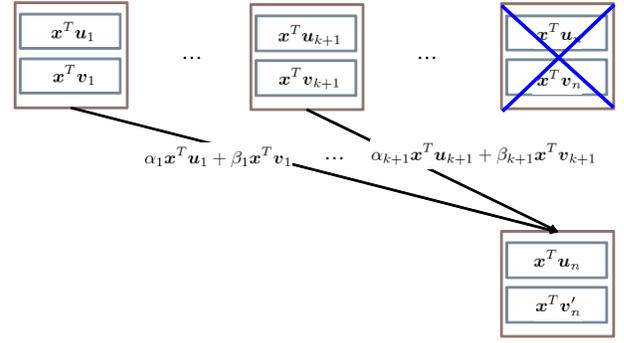

Fig. 2. Illustration of the proposed scheme.

set $P$ of $k-1$ other nodes in $\{2, \ldots, d+1\}$. Consider an $s$–$t$ cut $(U, \overline{U})$ with

$$\overline{U} \triangleq \{t, in_{n+1}, out_{n+1}\} \cup \{out_i : i \in P\}. \quad (1)$$

This is illustrated by Figure 1. Then we obtain a bound by requiring that the capacity of the cut is at least $B$

$$(k-1)\alpha + \sum_{i \notin P} \beta_i \geq B. \quad (2)$$

For each $(k-1)$-subset $P \subseteq \{2, \ldots, d+1\}$, we can obtain one inequality like (2). Summing up all these inequalities, we have that:

$$\binom{d}{k-1}\frac{d-k+1}{d}\sum_{i=1}^{d}\beta_i \geq \binom{d}{k-1}(B-(k-1)\alpha).$$

Thus

$$\sum_{i=1}^{d}\beta_i \geq \frac{Bd}{k(d-k+1)}. \qquad \blacksquare$$

## III. THE CODE CONSTRUCTION

The proposed scheme is illustrated in Figure 2. Let $\mathbb{F}$ denote the finite field where the code is defined in. In Figure 2, $\boldsymbol{x} \in \mathbb{F}^{2k}$ is a vector consisting of the $2k$ original information symbols. Each node stores 2 symbols, $\boldsymbol{x}^T \boldsymbol{u}_i$ and $\boldsymbol{x}^T \boldsymbol{v}_i$. The vectors $\{\boldsymbol{u}_i\}$ do not change over time. The vectors $\{\boldsymbol{v}_i\}$ changes over time as the code repairs. We maintain the invariant property that the $2n$ length-$2k$ vectors $\{\boldsymbol{u}_i, \boldsymbol{v}_i\}$ form an $(2n, 2k)$-MDS code; that is, any $2k$ vectors in the set $\{\boldsymbol{u}_i, \boldsymbol{v}_i\}$ has full rank $2k$. This certainly implies that the $n$ nodes form an $(n,k)$-MDS code. We initialize the code using any $(2n, 2k)$ systematic MDS code over $\mathbb{F}$.

Now we consider the situation of a repair. Without loss of generality, suppose node $n$ failed and is repaired by accessing nodes $1, \ldots, k+1$. As illustrated in Figure 2, the replacement node downloads $\alpha_i \boldsymbol{x}^T \boldsymbol{u}_i + \beta_i \boldsymbol{x}^T \boldsymbol{v}_i$ from each node of $\{1, \ldots, k+1\}$. Using these $k+1$ downloaded symbols, the replacement node computes two symbols $\boldsymbol{x}^T \boldsymbol{u}_n$ and $\boldsymbol{x}^T \boldsymbol{v}'_n$ as follows:

$$\sum_{i=1}^{k+1} (\alpha_i \boldsymbol{x}^T \boldsymbol{u}_i + \beta_i \boldsymbol{x}^T \boldsymbol{v}_i) = \boldsymbol{x}^T \boldsymbol{u}_n \quad (3)$$

$$\sum_{i=1}^{k+1} \rho_i (\alpha_i \boldsymbol{x}^T \boldsymbol{u}_i + \beta_i \boldsymbol{x}^T \boldsymbol{v}_i) = \boldsymbol{x}^T \boldsymbol{v}'_n \quad (4)$$

Note that $\boldsymbol{v}'_n$ is allowed to be different from $\boldsymbol{v}_n$; the property that we maintain is that the repaired code continues to be an $(2n, 2k)$-MDS code. Here $\{\alpha_i, \beta_i, \rho_i\}$ and $\boldsymbol{v}'_n$ are the variables that we can control. The following theorem shows that we can choose these variables so that (3) and (4) are satisfied and the repaired code continues to be an $(2n, 2k)$-MDS code.

**Theorem 1:** Let $\mathbb{F}$ be a finite field whose size is greater than

$$d_0 = 2 \begin{pmatrix} 2n-1 \\ 2k-1 \end{pmatrix}. \qquad (5)$$

Suppose the old code specified by $\{\boldsymbol{u}_i, \boldsymbol{v}_i\}$ is an $(2n, 2k)$-MDS code defined over $\mathbb{F}$. When node $n$ fails, there exists an assignment of the variables $\{\alpha_i, \beta_i, \rho_i\}$ such that (3) and (4) are satisfied and the repaired code continues to be an $(2n, 2k)$-MDS code.

**Proof:** We begin by examining the condition (3). Introduce

$$\boldsymbol{\eta} \triangleq [\alpha_1, \beta_1, \ldots, \alpha_{k+1}, \beta_{k+1}]^T \qquad (6)$$

$$\boldsymbol{A} \triangleq [\boldsymbol{u}_1, \boldsymbol{v}_1, \ldots, \boldsymbol{u}_{k+1}, \boldsymbol{v}_{k+1}]. \qquad (7)$$

Let $\eta_i$ denote the $i$-th entry of $\boldsymbol{\eta}$; let $\boldsymbol{a}_i$ denote the $i$-th column of $\boldsymbol{A}$. Then the condition (3) can be equivalently written in matrix form as:

$$\boldsymbol{A}\boldsymbol{\eta} = \boldsymbol{u}_n. \qquad (8)$$

Suppose two arbitrary entries of $\boldsymbol{\eta}$, say $\eta_i$ and $\eta_j$, are fixed at arbitrary given values. Let $\boldsymbol{\eta}_{\setminus \{i,j\}}$ denote the subvector of $\boldsymbol{\eta}$ after removing the $i$-th and $j$-th entry and $\boldsymbol{A}_{\setminus \{i,j\}}$ denote the submatrix of $\boldsymbol{A}$ after removing the $i$-th and $j$-th column. Since $\boldsymbol{u}_1, \ldots, \boldsymbol{u}_n, \boldsymbol{v}_1, \ldots, \boldsymbol{v}_n$ are code vectors of an $(2n, 2k)$ MDS code, any $2k$ columns of $\boldsymbol{A}$ have full rank $2k$; in particular, $\boldsymbol{A}_{\setminus \{i,j\}}$ is invertible. Then, to satisfy $\boldsymbol{A}\boldsymbol{\eta} = \boldsymbol{u}_n$ with given $\eta_i$ and $\eta_j$, $\boldsymbol{\eta}_{\setminus \{i,j\}}$ is uniquely determined as

$$\boldsymbol{\eta}_{\setminus \{i,j\}} = \boldsymbol{A}_{\setminus \{i,j\}}^{-1} [\boldsymbol{u}_n - \eta_i \boldsymbol{a}_i - \eta_j \boldsymbol{a}_j]. \qquad (9)$$

Thus, the solutions to (3) have two degrees of freedom, with exactly $\mathbb{F}^2$ solutions. Given any two entries of $\boldsymbol{\eta}$, say $\eta_i$ and $\eta_j$, there is a unique solution to (3) and the other entries are affine functions of $\eta_i$ and $\eta_j$. In particular, we can consider $\eta_1 = \alpha_1, \eta_2 = \beta_1$ as the two free parameters after considering (3).

After considering (3), we are left with $k+3$ degrees of freedom that we can tune. Let the variables $\{\alpha_1, \beta_1, \rho_1, \ldots, \rho_{k+1}\}$ be collectively represented by a vector $\boldsymbol{\xi}$ with $k+3$ entries in $\mathbb{F}$. Next we examine (4). From (4), $\boldsymbol{v}'_n$ is determined as:

$$\boldsymbol{v}'_n = \sum_{i=1}^{k+1} \rho_i (\alpha_i \boldsymbol{u}_i + \beta_i \boldsymbol{v}_i). \qquad (10)$$

It remains to prove that we can choose $\boldsymbol{\xi} \in \mathbb{F}^{k+3}$ such that the new code $\{\boldsymbol{u}_1, \ldots, \boldsymbol{u}_n, \boldsymbol{v}_1, \ldots, \boldsymbol{v}_{n-1}, \boldsymbol{v}'_n\}$ continues to be an $(2n, 2k)$-MDS code.

Since the old code $\{\boldsymbol{u}_1, \boldsymbol{v}_1, \ldots, \boldsymbol{u}_n, \boldsymbol{v}_n\}$ was an $(2n, 2k)$-MDS code, we just need to prove that $\boldsymbol{v}'_n$ can be made linearly independent of any $2k-1$ subset of $\mathcal{U} \triangleq \{\boldsymbol{u}_1, \ldots, \boldsymbol{u}_n, \boldsymbol{v}_1, \ldots, \boldsymbol{v}_{n-1}\}$. For any $2k-1$ subset $S$ of $\{1, \ldots, 2n-1\}$, let $\boldsymbol{U}_S$ denote the $2k \times (2k-1)$ matrix whose columns are given by the vectors in $\mathcal{U}$ indexed by $S$. Then the $(2n, 2k)$-MDS condition boils down to:

$$\prod_{S \subset \{1,\ldots,2n-1\},\, |S|=2k-1} \det\left([\boldsymbol{U}_S, \boldsymbol{v}'_n]\right) \neq 0. \qquad (11)$$

From (10) and the discussion above, we see that each entry of $\boldsymbol{v}'_n$ is a multivariate polynomial in $\boldsymbol{\xi}$. This implies that the left hand side of (11) can be viewed as a multivariate polynomial in $\boldsymbol{\xi}$; it can be shown that the total degree of this polynomial is at most $d_0$.

**Claim 1:** For any $S \subset \{1, \ldots, 2n-1\}$ with $|S| = 2k-1$, $\det\left([\boldsymbol{U}_S, \boldsymbol{v}'_n]\right) \neq 0$ for some $\boldsymbol{\xi} \in \mathbb{F}^{k+3}$.

**Proof of Claim:** The replacement node downloads one symbol each from nodes $1, \ldots, k+1$. Each node $i$ of $1, \ldots, k+1$ stores a pair of symbols $\boldsymbol{x}^T \boldsymbol{u}_i$ and $\boldsymbol{x}^T \boldsymbol{v}_i$. The matrix $\boldsymbol{U}_S$ has $2k-1$ columns; we also view it as a set of $2k-1$ column vectors. Thus there must exist a node, say $i^*$, in $1, \ldots, k+1$, satisfying either $\boldsymbol{u}_{i^*} \notin \boldsymbol{U}_S$ or $\boldsymbol{v}_{i^*} \notin \boldsymbol{U}_S$ or both.

Suppose $\boldsymbol{v}_{i^*} \notin \boldsymbol{U}_S$ for $i^* \in \{1, \ldots, k+1\}$. From the discussion earlier, given any two entries of $\boldsymbol{\eta}$, there is a unique solution to (3). In particular, we can let $\alpha_{i^*} = 0$ and $\beta_{i^*} = 1$; this maps uniquely to one assignment of $\alpha_1$ and $\beta_1$, according to (9). We further choose $\rho_{i^*} = 1$ and all other $\rho_i = 0$. With this choice of $\boldsymbol{\xi}$, $\boldsymbol{v}'_n = \boldsymbol{v}_{i^*}$. Since $\boldsymbol{v}_{i^*} \notin \boldsymbol{U}_S$ and the old code $\{\boldsymbol{u}_1, \boldsymbol{v}_1, \ldots, \boldsymbol{u}_n, \boldsymbol{v}_n\}$ was an $(2n, 2k)$-MDS code,

$$\det\left([\boldsymbol{U}_S, \boldsymbol{v}'_n]\right) = \det\left([\boldsymbol{U}_S, \boldsymbol{v}_{i^*}]\right) \neq 0, \qquad (12)$$

for this choice of $\boldsymbol{\xi}$.

The case $\boldsymbol{u}_{i^*} \notin \boldsymbol{U}_S$ follows similarly. ∎

Claim 1 implies that $\det\left([\boldsymbol{U}_S, \boldsymbol{v}'_n]\right)$ is a nonzero multivariate polynomial in $\boldsymbol{\xi}$, which further implies that the left hand side of (11) is a nonzero multivariate polynomial in $\boldsymbol{\xi}$. From the Schwartz–Zippel Theorem (quoted below as Lemma 2), for a finite field whose size is $|\mathbb{F}| > d_0$, there exists an assignment of $\boldsymbol{\xi} \in \mathbb{F}^{k+3}$ such that (11) holds. Thus the theorem follows. ∎

**Lemma 2 (Schwartz–Zippel Theorem (see, e.g., [7])):**
Let $Q(x_1, \ldots, x_n) \in \mathbb{F}[x_1, \ldots, x_n]$ be a multivariate polynomial of total degree $d_0$ (the total degree is the maximum degree of the additive terms and the degree of a term is the sum of exponents of the variables). Fix any finite set $\mathbb{S} \subseteq \mathbb{F}$, and let $r_1, \ldots, r_n$ be chosen independently and uniformly at random from $\mathbb{S}$. Then if $Q(x_1, \ldots, x_n)$ is not equal to a zero polynomial,

$$\Pr[Q(r_1, \ldots, r_n) = 0] \leq \frac{d_0}{|\mathbb{S}|}. \qquad (13)$$

**Corollary 1 (A Systematic $(n, k)$-MDS Code):**
The above scheme gives a construction of systematic $(n, k)$-MDS codes for $2k \leq n$ that achieves the minimum repair bandwidth when repairing from $k+1$ nodes.

**Proof:** Consider $n \geq 2k$. Note that in the above scheme, we can initialize the code $\{\boldsymbol{u}_1, \ldots, \boldsymbol{u}_n, \boldsymbol{v}_1, \ldots, \boldsymbol{v}_n\}$ with any $(2n, 2k)$-MDS code. In particular, we can use a systematic code and assign the $2k$ systematic code vectors to

$\{\boldsymbol{u}_1, \ldots, \boldsymbol{u}_{2k}\}$. Since $\{\boldsymbol{u}_1, \ldots, \boldsymbol{u}_n\}$ do not change over time, the code remains a systematic $(2n, 2k)$-MDS code. Thus the $n$ nodes form a systematic $(n, k)$-MDS code. The code repairs a failure by downloading $k+1$ symbols from $d = k+1$ nodes, with the total file size is $B = 2k$. This achieves the cut bound given in Lemma 1. ∎

### A. Code Construction Algorithm

From the proof of Theorem 1 and the Schwartz–Zippel Theorem, for a sufficiently large finite field $\mathbb{F}$, if we independently and uniformly draw each entry of $\boldsymbol{\xi}$, then (11) will hold with high probability. This can be used to develop a randomized code construction procedure.

We initialize the code using any $(2n, 2k)$ systematic MDS code over $\mathbb{F}$. Subsequently, for each repair, we randomly draw a vector $\boldsymbol{\xi}$ from $\mathbb{F}^{2k}$. For each drawing of $\boldsymbol{\xi}$, we compute the resulting $\boldsymbol{v}'_n$ and check whether it is linearly independent from the $\binom{2n-1}{2k-1}$ subsets of $\{\boldsymbol{u}_1, \ldots, \boldsymbol{u}_n, \boldsymbol{v}_1, \ldots, \boldsymbol{v}_{n-1}\}$ with cardinality $2k - 1$. The random drawing process can be repeated until the desired property is met.

### B. Structural Comparison with Other Schemes

The above code scheme has a simple structure. It starts with any given $(2n, 2k)$-MDS code, with $n$ code vectors exactly maintained. The other $n$ code vectors evolve over time as the code repairs. The invariant property is that the code is always a $(2n, 2k)$-MDS code (hence an $(n, k)$-MDS code). We now compare the structure of this code with other existing schemes.

Since the proposed scheme works for $d = k + 1$, we only consider the case $d = k + 1$ in the comparison. For $d = k+1$, all schemes to be discussed below store two symbols at each node, which are linear combinations of the $2k$ original information symbols. Thus all these schemes can be expressed in the same notation, where $\boldsymbol{x}$ denotes the $2k$ information symbols and node $i$ stores $\boldsymbol{x}^T \boldsymbol{u}_i$ and $\boldsymbol{x}^T \boldsymbol{v}_i$. However, the schemes differ in additional structural properties imposed to the code and also in how the repair is done.

First, the network coding scheme for the functional repair model [1]–[4] achieves the cut bound on total repair bandwidth. The code has a looser structure compared to the code proposed in this paper. In each repair, the two symbols can be repaired to two new symbols. The only requirement is that the $2n$ code vectors always form an $(n, k)$-MDS. However, in doing so, it is hard, if not impossible, to provide the systematic feature.

Second, the interference alignment scheme for the exact repair model [5] achieves the cut bound on total repair bandwidth for $k = 2$ but not for general $k$. The code is formed by two rows of $(n, k)$-MDS[1], each involving half of the variables. More precisely, let the original information symbols $\boldsymbol{x}$ be split into two vectors $\boldsymbol{y}$ and $\boldsymbol{z}$, each of length $k$. Then the first row of the code is a systematic $(n, k)$-MDS code applied to $\boldsymbol{y}$ and the second row is a different systematic $(n, k)$-MDS code applied to $\boldsymbol{z}$. The requirement is that each row is $(n, k)$-MDS and certain interference cancelation condition must be met. In this code, the first $k$ node store the $2k$ systematic symbols; in comparison, in the proposed code, the systematic symbols are spread in the first row, across the nodes.

Third, the scheme of [6] achieves the cut bound on total repair bandwidth. In this scheme, the $2k$ original information symbols are represented by two vectors $\boldsymbol{y}$ and $\boldsymbol{z}$, each of length $k$. Each node stores $\boldsymbol{y}^T \boldsymbol{u}_i$ and $\boldsymbol{y}^T \boldsymbol{v}_i + \boldsymbol{z}^T \boldsymbol{u}_i$. The first row of the code, given by the vectors $\{\boldsymbol{u}_i\}$, do not change over time and they can be any $(n, k)$-MDS code. The second row of the code is essentially the same code $\{\boldsymbol{u}_i\}$ applied to $\boldsymbol{z}$ plus a linear function of $\boldsymbol{y}$. Here the vectors $\{\boldsymbol{v}_i\}$ changes over time as the code repairs; they can assume arbitrary values.

---

[1] The code is viewed as a $2 \times n$ matrix (see Figure 2), where the columns correspond to the $n$ nodes.